# Diffusive Molecular Communication with Nanomachine Mobility

Neeraj Varshney and Aditya K. Jagannatham
Department of Electrical Engineering
Indian Institute of Technology Kanpur
Kanpur, India 208016
Email: {neerajv,adityaj}@iitk.ac.in

Pramod K. Varshney
Department of Electrical Engineering & Computer Science
Syracuse University
Syracuse, NY 13244, USA
Email: varshney@syr.edu

*Abstract*—This work presents a performance analysis for diffusive molecular communication with mobile transmit and receive nanomachines. To begin with, the optimal test is obtained for symbol detection at the receiver nanomachine. Subsequently, closed-form expressions are derived for the probabilities of detection and false alarm, probability of error, and capacity considering also aberrations such as multi-source interference, inter-symbol interference, and counting errors. Simulation results are presented to corroborate the theoretical results derived and also, to yield various insights into the performance of the system. Interestingly, it is shown that the performance of the mobile diffusive molecular communication can be significantly enhanced by allocating large fraction of total available molecules for transmission as the slot interval increases.

## I. INTRODUCTION

Molecular communication has attracted significant interest due to its applicability in novel biomedical, industrial, and surveillance applications [1], [2]. Efficient drug delivery and human body monitoring using communicating nano-robots, as described in [3]–[6], are a few examples of its unlimited potential. Since the potential applications are in the area of nano-medicine or nano-sensing, e.g., a healthcare application inside one of the blood vessels of a human body, it is relevant to consider the movement of nanomachines while analyzing the performance of the diffusive molecular communication systems. The development of such systems followed by their performance analysis is an active area in current research, which forms the focus of works [7]–[14]. Work in [7] presented an analytical framework to characterize the performance of electrochemical communication between nanomachines considering also the effects of mobility. The work in [8] proposed a maximum-Likelihood estimator (MLE) for the clock offset in a mobile molecular communication system for a scenario with one of the communicating nanomachines mobile. Authors in [9] proposed an adaptive code width protocol to mitigate the inter-symbol interference (ISI) with mobility employing feedback from the receiver nanomachine (RN) to transmitter nanomachine (TN). The work in [10] examined the impact of transposition errors on mobile molecular communication with positional-distance codes and demonstrated their improved performance over the traditional Hamming distance-based schemes. An analysis of the message delivery delay for a mobile nano-network has been presented in [11] considering bacteria as the message transmission medium. Authors in [12] proposed a molecular communication method based on Förster Resonance Energy Transfer (FRET) for mobile nanomachine networks and subsequently analyzed the probability of successful transmission, mean message extinction time, system throughput, channel capacity and achievable communication rates in such systems. Authors in [13] developed a model for non-diffusive mobile molecular communication networks. The performance of diffusive mobile molecular communication over time-varying channels has been analyzed in [14]. The recent work in [15] derives closed-form expression for the probability distribution of the first hitting time employing Brownian motion to model nanomachine mobility. In contrast to [14], the analysis therein incorporated the variation of the transmitter position towards derivation of the first hitting time distribution and subsequently presents an analysis for the resulting transposition error. However, the calculation of the expected bit error probability therein is only possible for small number of slots $K$ since the set of all possible permutation on $K$ released molecules has $K!$ elements. Thus, analysis of molecuelar communication considering mobile nanomachines considering realistic scenarios with multiple-source interference (MSI), inter-symbol interference (ISI), and counting errors is lacking in the existing literature.

This work, therefore, analyzes the performance of diffusive molecular communication with transmit and receive nanomachines that are mobile with Brownian motion, similar to [15]. However, in contrast to [15] and the references therein, analytical expressions are derived to characterize the probabilities of detection, false alarm, and error, and also the channel capacity incorporating practical distortions such as MSI, ISI, and counting errors at the receiver. Simulation results are presented that illustrate the performance and yield various insights into the performance of the system.

## II. SYSTEM MODEL

Consider a diffusive molecular communication system with both the transmitter and receiver nanomachines in motion while communicating in a semi-infinite one-dimensional (1D) fluid medium. This system model is similar to the one considered in the recent work [15] wherein a point source transmitter nanomachine and a point receiver nanomachine are placed on a straight line at a certain distance from each other. The movement of both nanoachines is modeled as a 1D Gaussian random walk. In addition, the channel is divided into time slots of duration $\tau$, where the $j$th slot is defined as the time period $[(j-1)\tau, j\tau]$ with $j \in \{1, 2, \cdots\}$. At the beginning of each time slot, the TN either emits the same type of molecules in the propagation medium with prior probability $\beta$ for transmission of information symbol 1 or remains silent for transmission of information symbol 0. Let $\mathcal{Q}[j]$ denote the number of molecules released by the TN for information symbol $x[j] = 1$ at the beginning of the $j$th slot. The molecular propagation from the TN to RN occurs via Brownian Motion with diffusion coefficient denoted by $D_p$. Similar to [16], [17] and the references therein, it is also assumed that the transmitted molecules do not interfere or collide with each other. Once the molecules reach the RN, they are immediately absorbed by it, followed by detection of the transmitted information based on the number of molecules received.

Due to the stochastic nature of the diffusive channel, the times of arrival at the RN, of the molecules emitted by the TN, are random in nature and can span multiple time slots. Let $q_{j-i}$ denote the probability that a molecule transmitted in slot $i \in \{1, 2, \cdots, j\}$ arrives in time slot $j$ and can be obtained as

$$q_{j-i} = \int_{(j-i)\tau}^{(j-i+1)\tau} f(t; i) dt, \quad (1)$$

where the first hitting time distribution $f(t; i)$ for a mobile TN and RN with diffusion coefficients $D_{\text{tx}}$ and $D_{\text{rx}}$ respectively, is obtained from [15, Eq. (6)] as

$$f(t;i) = \frac{\sqrt{i\tau D_{\text{tot}} D_{p,\text{eff}}}}{\pi \sqrt{t}(i\tau D_{\text{tot}} + t D_{p,\text{eff}})} \exp\left(-\frac{d^2}{4i\tau D_{\text{tot}}}\right)$$
$$+ \frac{d}{\sqrt{4\pi D_{p,\text{eff}}(t + i\tau D_{\text{tot}}/D_{p,\text{eff}})^3}}$$
$$\times \exp\left(-\frac{d^2}{4 D_{p,\text{eff}}(t + i\tau D_{\text{tot}}/D_{p,\text{eff}})}\right)$$
$$\times \text{erf}\left(\frac{d}{2}\sqrt{\frac{t D_{p,\text{eff}}}{i\tau D_{\text{tot}}(i\tau D_{\text{tot}} + t D_{p,\text{eff}})}}\right), \quad (2)$$

where $d$ is the Euclidean distance between the TN and RN, and $\text{erf}(x)$ denotes the standard error function [18]. The quantities $D_{\text{tot}}$ and $D_{p,\text{eff}}$ are defined as, $D_{\text{tot}} = D_{\text{tx}} + D_{\text{rx}}$ and $D_{p,\text{eff}} = D_{\text{rx}} + D_p$ respectively.

The number of molecules received at the RN during time slot $[(j-1)\tau, j\tau]$ can be expressed as

$$R[j] = S[j] + \mathcal{I}[j] + N[j] + C[j]. \quad (3)$$

The quantity $S[j]$ denotes the number of molecules received in the current slot $[(j-1)\tau, j\tau]$ and is binomially distributed with parameters $\mathcal{Q}[j]x[j]$ and $q_0$, where $x[j] \in \{0, 1\}$ is the symbol transmitted by the TN in the $j$th time slot. The quantity $N[j]$ denotes the MSI, i.e., noise arising due to molecules received from the other sources, which can be modeled as a Gaussian random variable with mean $\mu_o$ and variance $\sigma_o^2$ assuming a sufficiently large number of interfering sources [19]. Further, the noise $N[j]$ is independent of the number of molecules $S[j]$ and $\mathcal{I}[j]$ received from the intended TN [16]. The term $C[j]$ denotes counting error at the RN and can be modeled as a Gaussian distributed random variable with zero mean and variance $\sigma_c^2[j]$ that depends on the average number of molecules received, i.e., $\sigma_c^2[j] = \mathbb{E}\{R[j]\}$ [16], [20]. The quantity $\mathcal{I}[j]$ models the ISI, arising from transmission in the previous $j-1$ slots, and is determined as

$$\mathcal{I}[j] = I[1] + I[2] + \cdots + I[j-1], \quad (4)$$

where $I[i] \sim Binomial(\mathcal{Q}[j-i]x[j-i], q_i), 1 \leq i \leq j-1$, denotes the number of stray molecules received from the previous $(j-i)$th slot. Assuming the number of molecules released by TN to be sufficiently large, the binomial distribution for $S[j]$ can be approximated by a Gaussian distribution[1] with mean $\mu[j] = \mathcal{Q}[j]x[j]q_0$ and variance $\sigma^2[j] = \mathcal{Q}[j]x[j]q_0(1-q_0)$, i.e., $S[j] \sim \mathcal{N}(\mathcal{Q}[j]x[j]q_0, \mathcal{Q}[j]x[j]q_0(1-q_0))$ [21]. Similarly, the binomial distribution for $I[i], 1 \leq i \leq j-1$ can be approximated as $I[i] \sim \mathcal{N}(\mu_I[i] = \mathcal{Q}[j-i]x[j-i]q_i, \sigma_I^2[i] = \mathcal{Q}[j-i]x[j-i]q_i(1-q_i))$. Further, it can be noted that $S[j]$ and $I[i], i = 1, 2, \cdots, j-1$ are mutually independent since the molecules transmitted in different time slots do not interfere with each other [16], [17].

## III. Performance Analysis

The symbol detection problem at the RN can be formulated as the binary hypothesis testing problem

$$\begin{aligned}\mathcal{H}_0 &: R[j] = I[1] + I[2] + \cdots + I[j-1] + N[j] + C[j] \\ \mathcal{H}_1 &: R[j] = S[j] + I[1] + I[2] + \cdots + I[j-1] \\ &\quad + N[j] + C[j],\end{aligned} \quad (5)$$

where the null and alternative hypotheses $\mathcal{H}_0$, $\mathcal{H}_1$ correspond to the transmission of binary symbols 0, 1 respectively during the $j$th time slot. The number of molecules $R[j]$ received at the RN corresponding to the different hypotheses are distributed as

$$\begin{aligned}\mathcal{H}_0 &: R[j] \sim \mathcal{N}(\mu_0[j], \sigma_0^2[j]) \\ \mathcal{H}_1 &: R[j] \sim \mathcal{N}(\mu_1[j], \sigma_1^2[j]),\end{aligned} \quad (6)$$

where the mean $\mu_0[j]$ and variance $\sigma_0^2[j]$ under hypothesis $\mathcal{H}_0$, as derived in Appendix A, is given as

$$\mu_0[j] = \mu_I[1] + \mu_I[2] + \cdots + \mu_I[j-1] + \mu_o$$
$$= \beta \sum_{i=1}^{j-1} \mathcal{Q}[j-i]q_i + \mu_o, \quad (7)$$
$$\sigma_0^2[j] = \sigma_I^2[1] + \sigma_I^2[2] + \cdots + \sigma_I^2[j-1] + \sigma_o^2 + \sigma_c^2[j]$$
$$= \sum_{i=1}^{j-1} [\beta \mathcal{Q}[j-i]q_i(1-q_i) + \beta(1-\beta)$$
$$\times (\mathcal{Q}[j-i]q_i)^2] + \sigma_o^2 + \mu_0[j], \quad (8)$$

[1] This approximation is reasonable when $\mathcal{Q}[j]q_0 > 5$ and $\mathcal{Q}[j](1-q_0) > 5$ [16].

and the mean $\mu_1[j]$ and variance $\sigma_1^2[j]$ under hypothesis $\mathcal{H}_1$, as derived in Appendix B, is given as

$$\mu_1[j] = \mu[j] + \mu_I[1] + \mu_I[2] + \cdots + \mu_I[j-1] + \mu_o$$
$$= \mathcal{Q}[j]q_0 + \beta \sum_{i=1}^{j-1} \mathcal{Q}[j-i]q_i + \mu_o, \quad (9)$$

$$\sigma_1^2[j] = \sigma^2[j] + \sigma_I^2[1] + \sigma_I^2[2] + \cdots + \sigma_I^2[j-1] + \sigma_o^2 + \sigma_c^2[j]$$
$$= \mathcal{Q}[j]q_0(1-q_0) + \sum_{i=1}^{j-1} [\beta \mathcal{Q}[j-i]q_i(1-q_i)$$
$$+ \beta(1-\beta)(\mathcal{Q}[j-i]q_i)^2] + \sigma_o^2 + \mu_1[j]. \quad (10)$$

The next result derives the optimal decision rule at the RN for symbol detection.

*Theorem 1:* The optimal decision rule at the RN corresponding to the $j$th time slot is obtained as

$$T(R[j]) = R[j] \underset{\mathcal{H}_0}{\overset{\mathcal{H}_1}{\gtrless}} \gamma'[j], \quad (11)$$

where the optimal decision threshold $\gamma'[j]$ is given as

$$\gamma'[j] = \sqrt{\gamma[j]} - \alpha[j]. \quad (12)$$

The quantities $\gamma[j]$ and $\alpha[j]$ are defined as

$$\alpha[j] = \frac{\mu_1[j]\sigma_0^2[j] - \mu_0[j]\sigma_1^2[j]}{\sigma_1^2[j] - \sigma_0^2[j]}, \quad (13)$$

$$\gamma[j] = \frac{2\sigma_1^2[j]\sigma_0^2[j]}{\sigma_1^2[j] - \sigma_0^2[j]} \ln\left[\frac{(1-\beta)}{\beta}\sqrt{\frac{\sigma_1^2[j]}{\sigma_0^2[j]}}\right] + (\alpha[j])^2$$
$$+ \frac{\mu_1^2[j]\sigma_0^2[j] - \mu_0^2[j]\sigma_1^2[j]}{\sigma_1^2[j] - \sigma_0^2[j]}. \quad (14)$$

**Proof** The optimal log likelihood ratio test (LLRT) at the RN is given as

$$\Lambda(R[j]) = \ln\left[\frac{p(R[j]|\mathcal{H}_1)}{p(R[j]|\mathcal{H}_0)}\right] \underset{\mathcal{H}_0}{\overset{\mathcal{H}_1}{\gtrless}} \ln\left[\frac{1-\beta}{\beta}\right]. \quad (15)$$

Substituting the Gaussian PDFs $p(R[j]|\mathcal{H}_1)$ and $p(R[j]|\mathcal{H}_0)$ from (6), the test statistic $\Lambda(R[j])$ can be obtained as

$$\Lambda(R[j]) = \ln\left[\sqrt{\frac{\sigma_0^2[j]}{\sigma_1^2[j]}}\right] + \frac{1}{2\sigma_0^2[j]\sigma_1^2[j]}$$
$$\times \underbrace{(R[j]-\mu_0[j])^2\sigma_1^2[j] - (R[j]-\mu_1[j])^2\sigma_0^2[j]}_{\triangleq f(R[j])}. \quad (16)$$

The expression for $f(R[j])$ given above can be further simplified as

$$f(R[j]) = R^2[j](\sigma_1^2[j] - \sigma_0^2[j]) + 2R[j](\mu_1[j]\sigma_0^2[j]$$
$$- \mu_0[j]\sigma_1^2[j]) + (\mu_0^2[j]\sigma_1^2[j] - \mu_1^2[j]\sigma_0^2[j])$$
$$= (\sigma_1^2[j] - \sigma_0^2[j])(R[j] + \alpha[j])^2$$
$$- \frac{(\mu_1[j]\sigma_0^2[j] - \mu_0[j]\sigma_1^2[j])^2}{\sigma_1^2[j] - \sigma_0^2[j]}$$
$$+ (\mu_0^2[j]\sigma_1^2[j] - \mu_1^2[j]\sigma_0^2[j]), \quad (17)$$

where $\alpha[j]$ is defined in (13). Substituting the above equation for $f(R[j])$ in (16) and subsequently merging the terms independent of the received molecules $R[j]$ with the detection threshold, the test can be equivalently expressed as

$$(R[j] + \alpha[j])^2 \underset{\mathcal{H}_0}{\overset{\mathcal{H}_1}{\gtrless}} \gamma[j], \quad (18)$$

where $\gamma[j]$ is defined in (14). Further, taking the square root of both sides where $\gamma[j] \geq 0$, Equation (18) can be simplified to yield the optimal test in (11). ∎

*A. Detection Performance Analysis*

The detection performance for the optimal test derived in (11) at RN considering also mobility of TN and RN, is obtained next.

*Theorem 2:* The average probabilities of detection $(P_D)$ and false alarm $(P_{FA})$ at the RN in the diffusion based mobile molecular communication nano-network, corresponding to the transmission by the TN over $k$ slots, are given as

$$P_D = \frac{1}{k}\sum_{j=1}^{k} P_D[j]$$
$$= \frac{1}{k}\sum_{j=1}^{k} Q\left(\frac{\gamma'[j] - \mu_1[j]}{\sigma_1[j]}\right), \quad (19)$$

$$P_{FA} = \frac{1}{k}\sum_{j=1}^{k} P_{FA}[j]$$
$$= \frac{1}{k}\sum_{j=1}^{k} Q\left(\frac{\gamma'[j] - \mu_0[j]}{\sigma_0[j]}\right), \quad (20)$$

where $\gamma'[j]$ is defined in (12) and $Q(\cdot)$ denotes the tail probability of the standard normal random variable [22].

**Proof** The probabilities of detection $(P_D[j])$ and false alarm $(P_{FA}[j])$ at the RN in the $j$th time slot for the decision rule in (11) are obtained as

$$P_D[j] = \Pr(T(R[j]) > \gamma'[j]|\mathcal{H}_1)$$
$$= \Pr(R[j] > \gamma'[j]|\mathcal{H}_1), \quad (21)$$
$$P_{FA}[j] = \Pr(T(R[j]) > \gamma'[j]|\mathcal{H}_0)$$
$$= \Pr(R[j] > \gamma'[j]|\mathcal{H}_0), \quad (22)$$

where number of received molecules $R[j]$ is Gaussian distributed (6) under hypotheses $\mathcal{H}_0$ and $\mathcal{H}_1$ respectively. Subtracting their respective means followed by division by the standard deviations, i.e., $\frac{R[j]-\mu_1[j]}{\sigma_1[j]}$ and $\frac{R[j]-\mu_0[j]}{\sigma_0[j]}$ yields standard normal random variables for hypotheses $\mathcal{H}_1$ and $\mathcal{H}_0$ respectively. Subsequently, the expressions for $P_D[j]$ and $P_{FA}[j]$, given in (19) and (20) respectively, can be obtained employing the definition of the $Q(\cdot)$ function. ∎

*B. Probability of Error Analysis*

The end-to-end probability of error for communication between TN and RN follows as described in the result below.

*Lemma 1:* The average probability of error ($P_e$) for slots 1 to $k$ at the RN in the diffusion based molecular nano-network with mobile TN and RN is

$$P_e = \frac{1}{k}\sum_{j=1}^{k}\left[\beta\left(1 - Q\left(\frac{\gamma'[j]-\mu_1[j]}{\sigma_1[j]}\right)\right) \right.$$
$$\left. + (1-\beta)Q\left(\frac{\gamma'[j]-\mu_0[j]}{\sigma_0[j]}\right)\right]. \quad (23)$$

**Proof** The probability of error $P_e[j]$ in $j$th time slot is defined as [22]

$$P_e[j] = \Pr(\text{decide }\mathcal{H}_0, \mathcal{H}_1\text{ true}) + \Pr(\text{decide }\mathcal{H}_1, \mathcal{H}_0\text{ true})$$
$$= (1 - P_D[j])P(\mathcal{H}_1) + P_{FA}[j]P(\mathcal{H}_0), \quad (24)$$

where the prior probabilities of the hypotheses $P(\mathcal{H}_1)$ and $P(\mathcal{H}_0)$ are $\beta$ and $1-\beta$ respectively. The quantities $P_D[j]$ and $P_{FA}[j]$ denote the probabilities of detection and false alarm at RN during the $j$th time slot as obtained in (19) and (20) respectively. The average probability of error for slots 1 to $k$ follows as stated in (23). ■

### C. Capacity Analysis

Let $X[j]$ and $Y[j]$ be two discrete random variables that represent the transmitted and received symbols, respectively, in the $j$th slot. The mutual information $I(X[j], Y[j])$ between $X[j]$ and $Y[j]$ with marginal probabilities $\Pr(x[j]=0) = 1-\beta, \Pr(x[j]=1) = \beta$ is given by

$$I(X[j],Y[j]) = \sum_{x[j]\in\{0,1\}}\sum_{y[j]\in\{0,1\}}\Pr(y[j]|x[j])\Pr(x[j])$$
$$\times \log_2 \frac{\Pr(y[j]|x[j])}{\sum_{x[j]\in\{0,1\}}\Pr(y[j]|x[j])\Pr(x[j])}, \quad (25)$$

where the conditional probabilities $\Pr(y[j] \in \{0,1\}|x[j] \in \{0,1\})$ can be written in terms of $P_D[j]$ and $P_{FA}[j]$ as

$$\Pr(y[j]=0|x[j]=0) = 1 - P_{FA}[j]$$
$$= 1 - Q\left(\frac{\gamma'[j]-\mu_0[j]}{\sigma_0[j]}\right),$$
$$\Pr(y[j]=1|x[j]=0) = P_{FA}[j]$$
$$= Q\left(\frac{\gamma'[j]-\mu_0[j]}{\sigma_0[j]}\right),$$
$$\Pr(y[j]=0|x[j]=1) = 1 - P_D[j]$$
$$= 1 - Q\left(\frac{\gamma'[j]-\mu_1[j]}{\sigma_1[j]}\right),$$
$$\Pr(y[j]=1|x[j]=1) = P_D[j]$$
$$= Q\left(\frac{\gamma'[j]-\mu_1[j]}{\sigma_1[j]}\right).$$

The mutual information between the TN and RN can be maximized as

$$C[k] = \max_{\beta}\frac{1}{k}\sum_{j=1}^{k}I(X[j],Y[j]) \text{ bits/slot}, \quad (26)$$

which equals the channel capacity as the number of slots $k \to \infty$ [23].

## IV. SIMULATION RESULTS

This section presents simulation results to demonstrate the performance of diffusive mobile molecular communication system under consideration. For simulation purposes, the various parameters are set as, diffusion coefficient $D_p = 5\times 10^{-10}$ m$^2$/s, slot duration $\tau = 0.01$ms, distance $d = 1$ $\mu$m, prior probability $\beta = 0.5$, and $k = 20$ slots. The MSI at each receiving node is modeled as a Gaussian distributed RV with mean $\mu_o = 10$ and variance $\sigma_o^2 = 10$ unless otherwise stated.

Fig. 1 shows that the analytical values obtained for probabilities of detection and false alarm, probability of error and capacity using (19), (20), (23), and (26) respectively coincide with those obtained from simulations, thus validating the analysis. Fig. 1(a) demonstrates the detection performance at the RN for various scenarios. It can be clearly seen that an increase in the number of molecules emitted by the TN results in a higher probability of detection at the RN for a fixed value of probability of false alarm. However, the detection performance significantly deteriorates as the diffusion coefficients $D_{\text{tx}}$ and $D_{\text{rx}}$ increase due to higher mobility. Fig. 1(b) presents the error rate versus noise variance $\sigma_o^2$ performance for several values of $D_{\text{tx}}$ and $D_{\text{rx}}$. One can observe that an increase in the noise variance $\sigma_o^2$ results in a higher probability of error at the RN. Moreover, the error rate further increases as the mobility increases. Fig. 1(c) shows the capacity performance where the maximum mutual information is achieved for equiprobable information symbols, i.e., $\beta = 0.5$. Similar to the error rate performance, the maximum mutual information $C[k]$ obtained for slots 1 to $k$ also decreases as the noise variance ($\sigma_o^2$) increases. Interestingly, one can also observe that the maximum mutual information $C[k]$ in bits per channel use progressively decreases as the number of slots ($k$) increases due to the mobile nature of TN and RN. This is owing to the fact that the probability of a molecule reaching RN within the current slot, i.e., $q_0$ progressively decreases while the ISI from previous slots increases as the value of $k$ increases.

Further, to demonstrate the impact of the number of molecules, we consider transmission during the first two time slots. Fig. 2 shows the probability of error versus the number of molecules $\mathcal{Q}[1]$ transmitted in slot $j = 1$ for various mobile and MSI noise scenarios, for a fixed molecule budget i.e., $\mathcal{Q}[1] + \mathcal{Q}[2] = 60$. Firstly, it can be observed from Figs. 2(a)-(b) that allocating equal number of molecules, $\mathcal{Q}[1] = \mathcal{Q}[2] = 30$, in the first and second time slots is not optimal in all the scenarios. Second, the performance of the diffusive molecular communication can be significantly enhanced by allocating large fraction of total available molecules in the second time slot in comparison to the first time slot as the diffusion coefficients $D_{\text{tx}}, D_{\text{rx}}$ increase due to higher mobility. This arises due to the progressive reduction in $q_0$ from 0.4505 to 0.3480 and 0.7511 to 0.7338 as the slot interval increases for $D_{\text{tx}} = D_{\text{rx}} = 10^{-9}$ m$^2$/s and $D_{\text{tx}} = D_{\text{rx}} = 10^{-11}$ m$^2$/s respectively. Further, one can also observe from Fig. 2(b) that the optimal number of molecules $\mathcal{Q}[1]$ transmitted in the first time slot increases from 25 to 28 as the MSI noise variance $\sigma_o^2$ increases from 1 to 20. This leads to the conclusion that the optimal molecular allocation tends towards equal distribution across slots as the MSI increases.

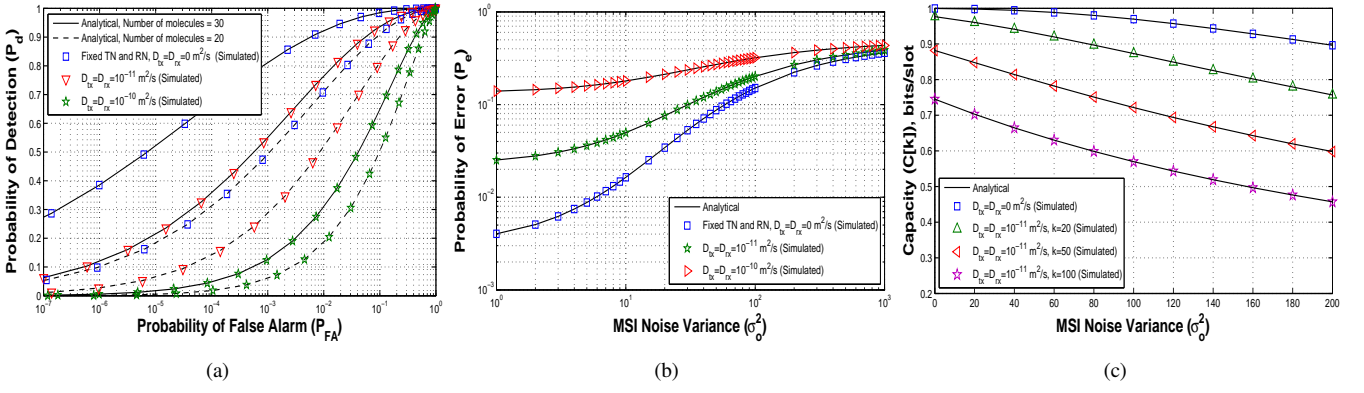

Fig. 1: With mobility: (a) Probability of detection versus probability of false alarm with number of molecules $\mathcal{Q}[j] \in \{20, 30\}, \forall j$. (b) Probability of error versus MSI noise variance ($\sigma_o^2$) with $\mu_o = 0$ and $\mathcal{Q}[j]=30, \forall j$. (c) Capacity versus $\sigma_o^2$ with $\mu_o = 10$ and $\mathcal{Q}[j]=100, \forall j$.

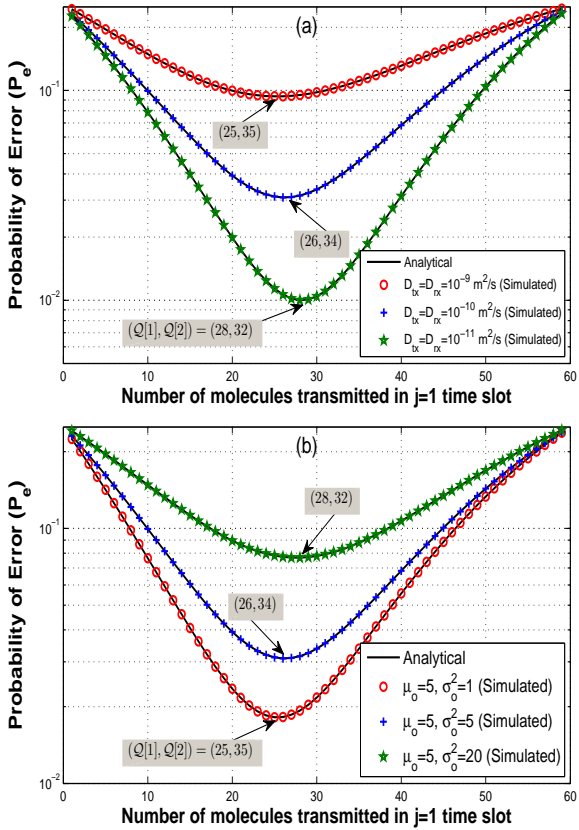

Fig. 2: With mobility: (a) Probability of error versus number of molecules $\mathcal{Q}[1]$ transmitted in $j = 1$ slot for varying $D_{\text{tx}}$, $D_{\text{rx}}$ and fixed MSI with $\mu_o = \sigma_o^2 = 5$. (b) Probability of error versus number of molecules $\mathcal{Q}[1]$ transmitted in $j = 1$ slot for varying MSI noise variance $\sigma_o^2$ and fixed mobility with $D_{\text{tx}} = D_{\text{rx}} = 10^{-10}$ m$^2$/s.

## V. CONCLUSION

This work analyzed the performance of diffusive molecular communication considering TN and RN mobility together with MSI, ISI, and counting errors. The optimal decision rule was obtained followed by analytical results for the probabilities of detection, false alarm, error and channel capacity. It was observed that the probability of error and capacity degrade significantly due to mobile nature of the TN and RN. Further, performance was seen to improve significantly by optimally allocating the molecules over time slots with a fixed molecule budget. Finally, future studies can focus on the optimization framework to allocate the transmitted molecules at each time-slot to enhance the performance of diffusive mobile molecular communication.

## APPENDIX A
### MEAN $\mu_0[j]$ AND VARIANCE $\sigma_0^2[j]$ UNDER HYPOTHESIS $\mathcal{H}_0$

Using (5), the mean $\mu_0[j]$ under $\mathcal{H}_0$ can be calculated as

$$\mu_0[j] = \mathbb{E}\left\{\sum_{i=1}^{j-1} I[i] + N[j] + C[j]\right\}$$
$$= \sum_{i=1}^{j-1} \mathbb{E}\{I[i]\} + \mu_o, \quad (27)$$

where $\mathbb{E}\{I[i]\}$ is given as

$$\mathbb{E}\{I[i]\} = \Pr(x[j-i]=1)\mathbb{E}\{I[i]|x[j-i]=1\}$$
$$+ \Pr(x[j-i]=0)\mathbb{E}\{I[i]|x[j-i]=0\}$$
$$= \beta\mathbb{E}\{I[i]|x[j-i]=1\} + (1-\beta)\mathbb{E}\{I[i]|x[j-i]=0\}$$
$$= \beta\mathcal{Q}[j-i]q_i. \quad (28)$$

The variance $\sigma_0^2[j]$ under $\mathcal{H}_0$ can be derived as

$$\sigma_0^2[j] = \sum_{i=1}^{j-1} \sigma_I^2[i] + \sigma_o^2 + \sigma_c^2[j]$$
$$= \sum_{i=1}^{j-1} \sigma_I^2[i] + \sigma_o^2 + \mu_0[j], \quad (29)$$

where $\sigma_c^2[j] = \mu_0[j]$ and the variance $\sigma_I^2[i]$ of ISI term can be obtained as

$$\sigma_I^2[i] = \mathbb{E}\{(I[i])^2\} - \mathbb{E}^2\{I[i]\} \\ = \mathbb{E}\{(I[i])^2\} - (\beta \mathcal{Q}[j-i]q_i)^2, \quad (30)$$

where $\mathbb{E}\{(I[i])^2\}$ is given as

$$\mathbb{E}\{(I[i])^2\} = \Pr(x[j-i]=1)\mathbb{E}\{(I[i])^2|x[j-i]=1\} \\ + \Pr(x[j-i]=0)\mathbb{E}\{(I[i])^2|x[j-i]=0\} \\ = \beta \mathbb{E}\{(I[i])^2|x[j-i]=1\} \\ = \beta \left[\mathcal{Q}[j-i]q_i(1-q_i) + (\mathcal{Q}[j-i]q_i)^2\right]. \quad (31)$$

Substituting the above expression in (30), the final expression for the variance $\sigma_I^2[i]$ of ISI term is given as

$$\sigma_I^2[i] = \beta \mathcal{Q}[j-i]q_i(1-q_i) + \beta(1-\beta)(\mathcal{Q}[j-i]q_i)^2. \quad (32)$$

## APPENDIX B
MEAN $\mu_1[j]$ AND VARIANCE $\sigma_1^2[j]$ UNDER HYPOTHESIS $\mathcal{H}_1$

Similar to $\mu_0[j]$, the mean $\mu_1[j]$ under $\mathcal{H}_1$ can be calculated using (5) as

$$\mu_1[j] = \mathbb{E}\left\{S[j] + \sum_{i=1}^{j-1} I[i] + N[j] + C[j]\right\} \\ = \mathbb{E}\{S[j]\} + \sum_{i=1}^{j-1} \mathbb{E}\{I[i]\} + \mu_o \\ = \mathcal{Q}[j]q_0 + \beta \sum_{i=1}^{j-1} \mathcal{Q}[j-i]q_i + \mu_0. \quad (33)$$

The variance $\sigma_1^2[j]$ under $\mathcal{H}_1$ can be derived as

$$\sigma_1^2[j] = \sigma^2[j] + \sum_{i=1}^{j-1} \sigma_I^2[i] + \sigma_o^2 + \sigma_c^2[j] \\ = \sigma^2[j] + \sum_{i=1}^{j-1} \sigma_I^2[i] + \sigma_o^2 + \mu_1[j], \quad (34)$$

where $\sigma^2[j] = \mathcal{Q}[j]q_0(1-q_0)$, $\sigma_c^2[j] = \mu_1[j]$, and $\sigma_I^2[i]$ is given in (32).